\documentclass[lettersize,journal]{IEEEtran}
\usepackage{amsmath,amsfonts}
\usepackage{algorithmic}
\usepackage{array}
\usepackage{subcaption} 
\usepackage{textcomp}
\usepackage{stfloats}
\usepackage{url}
\usepackage{verbatim}
\usepackage{graphicx}
\def\BibTeX{{\rm B\kern-.05em{\sc i\kern-.025em b}\kern-.08em
    T\kern-.1667em\lower.7ex\hbox{E}\kern-.125emX}}
\usepackage{balance}
\usepackage{color}
\RequirePackage{hyperref}
\usepackage[utf8]{inputenc}
\usepackage{pythonhighlight}
\usepackage{multirow} 

\begin{document}
\title{Palace: A Library for Interactive GPU-Accelerated Large Tensor Processing and Visualization}
\author{Dominik Drees and Benjamin Risse}


\maketitle

\begin{abstract}
  Tensor datasets (two-, three-, or higher-dimensional) are fundamental to many scientific fields utilizing imaging or simulation technologies.
  Advances in these methods have led to ever-increasing data sizes and, consequently, interest and development of out-of-core processing and visualization techniques, although mostly as specialized solutions.
  Here we present Palace, an open-source, cross-platform, general-purpose library for interactive and accelerated out-of-core tensor processing and visualization.
  Through a high-performance asynchronous concurrent architecture and a simple compute-graph interface, Palace enables the interactive development of out-of-core pipelines on workstation hardware.
  We demonstrate on benchmarks that Palace outperforms or matches state-of-the-art systems for volume rendering and hierarchical random-walker segmentation and demonstrate applicability in use cases involving tensors from 2D images up to 4D time series datasets.
\end{abstract}


\section{Introduction}
\IEEEPARstart{T}{ensor} data sets of varying dimensionality have increased in size in recent years, outgrowing the VRAM and potentially RAM capacity of workstation PCs.
This is true for 2D~\cite{kockwelp2024deep} and 3D~\cite{kirschnick2021rapid} data, particularly in the biomedical domain, but also for higher-dimensional data generated from simulations~\cite{kruger2017lattice}.
The problem has been a topic of interest in the visualization research community~\cite{bankhead2017qupath,sarton2023review}, but has also been noted in the context of medical image processing~\cite{eklund2013processinggpu}, where the need for GPU-based processing (given the enormous data sizes and associated compute times) has simultaneously been stated.
Despite the interest in the topic of out-of-core visualization and processing, most research produces specialized solutions for a specific narrowly defined problem (e.g., 2D~\cite{bankhead2017qupath} and 3D~\cite{drees2022voreen} visualization or specific processing algorithms~\cite{drees2021graph}).
A general framework enabling the development of out-of-core pipelines and algorithms has not yet been proposed.


In this work, we present Palace, a user-friendly library for interactive, GPU-accelerated out-of-core tensor processing and visualization.
The name Palace is an acronym for \textbf{p}rogressive \textbf{a}ccelerated \textbf{l}arge \textbf{a}rray \textbf{c}omputing \textbf{e}ngine.
With Palace, we aim to enable the development of processing and rendering pipelines for out-of-core tensor data on workstations rather than compute clusters), fully utilizing the available hardware, enabling use of all GPUs in the system and efficient pipelines that are able to saturate modern SSD bandwidth.
This supports fast prototyping via fast visualization of results for the given processing parameters using level-of-detail (LOD) data representations for a result overview, and a chunked pull-based architecture for selective computation of visualized high-resolution parts.
The dimension-agnostic implementation of most operators enables application in multi-modal scenarios. 

Palace is cross-platform, both in terms of the operating system (it contains little platform-specific code and supports Unix-like systems and Microsoft Windows) and the GPU vendor (by using the Vulkan~\cite{khronos2025vulkan} API for compute and rendering).
Palace is free software and available online\footnote{\url{https://github.com/ftilde/palace}} under the Mozilla Public License (MPL 2.0).

In the following sections, we first discuss related work and then detail Palace's architecture.
Afterward, we compare Palace's performance with state-of-the-art systems in scenarios such as volume rendering and tensor processing.
We also demonstrate Palace's applicability to other dimensionalities with examples




\section{Related Work}

Handling of larger-than-main-memory tensor data has been a recurring topic in the scientific community over the past two decades.
One particular area of interest lies in the visualization of 3D-tensor, i.e.,\ volume data.
In the early 2000s, LaMar~\cite{lamar2000multiresolution} and Weiler~\cite{weiler2000level} proposed out-of-core rendering algorithms for volumetric data based on voxel octrees, which has been used for visualization~\cite{crassin2009gigavoxels,brix2014visualization} and processing~\cite{jia2020using,drees2022octree} in the following years.

A further innovation was the introduction of the multi-resolution page directory for out-of-core rendering~\cite{hadwiger2012pth} which avoids the need to traverse the octree structure from the root node.
It borrows the idea of page table hierarchies of virtual memory~\cite{hennessy2011computer}, constructing a shallow, sparse tree with volume bricks as the leaves.
This work by Hadwiger et al.\ is notable for its on-demand brick and LOD (level-of-detail) construction and its CPU/GPU cache system, and it is tailored to the demands of microscopy data visualization.

Sarton et al.~\cite{sarton2020gpuooc} also use a multi-resolution page directory to access the volume's LOD pyramid.
It features an LRU-based VRAM cache for bricks and page-table pages, compressed brick storage, offline LOD construction, and limited in-place processing prior to visualization.
Notably, page table hierarchy updates are executed sequentially and requests are handled via one-entry-per-brick buffers, the latter limiting scaling to very large volumes (without increasing the brick size).

In addition, several recent works have focused on out-of-core computation and visualization~\cite{wu2022streamingvis,sarton2019distributed,cousty2021algebraic}."
A general overview of state-of-the art large volume visualization can be found in a review~\cite{sarton2023review} by Sarton et al.

For higher-dimensional tensor data, out-of-core processing and visualization approaches are in high demand, due to ever larger data sizes as a result of four or more dimensions~\cite{barff2023timeseries, panta2024web}.

However, even for 2D images, as a result of increasing photographic resolution and mosaicing techniques, with applications in biomedical image analysis, data sizes frequently reach or exceed the memory capacity of available systems \cite{ha2012optimal,bankhead2017qupath}.

Notable software of recent years includes Terafly~\cite{bria2015terafly}, a plugin for V3D~\cite{peng2010v3d}, which uses a block-based approach with LOD layers and includes a slice viewer.
Ashwini et al.~\cite{ashwini2018teraweb} use Terafly to build an application for mouse brain scans of knife-edge scanning microscopy.
Miettinen et al.~\cite{miettinen2019nrstitcher} present a stitcher for teravoxel images that enables out-of-core processing by selectively reading the relevant parts of volumes.

Of further note is the software Dask~\cite{rocklin2015dask}, which (like Palace) enables chunked, pull-based processing of large tensor data via execution of a user-defined compute graph.
While Palace targets low-latency interactive processing and visualization on workstations, Dask's primary deployment is for exabyte-scale datasets on compute clusters.
Nevertheless, local execution is still possible: Napari~\cite{chiu2022napari} uses Dask and vispy~\cite{campagnola2015vispy} for visualization of biomedical volume data, but is limited to rendering volumes that fit into graphics memory (at least for raycasting). 

Palace combines Dask-like user-controlled computation graphs and chunked, pull-based execution with state-of-the-art GPU memory management and visualization techniques to facilitate rapid development of out-of-core tensor-processing pipelines.

\section{Architecture}

The following sections describe the architecture of Palace, starting from a user's view and then moving to the internal architectural details that enable that view.

\subsection{User's View}

The key concepts of Palace are tensor chunking and the pull-based architecture.
Chunking refers to the rectangular partitioning of tensors into equally sized \textit{chunks}, which are stored separately and enable partial loading of tensors into memory.
Given a $d$-dimensional tensor with size $S \in \mathbb{N}_+^d$ and chunk size $C \in \mathbb{N}_+^d$, a global position $g \in \mathbb{N}_+^d$ can be converted into a chunk position $h\in \mathbb{N}_+^d$ and local position $l\in \mathbb{N}_+^d$ via division and modulo operations:
\begin{align}
  h_i &= \lfloor g_i / C_i \rfloor\nonumber\\
  l_i &= g_i \bmod C_i\nonumber
\end{align}
The chunk position $h$ can then be used to locate the respective chunk, and the local position $l$ is used to find the element value within a chunk.
Two-dimensional chunks are also called \textit{tiles}, and three-dimensional chunks \textit{bricks}.

In the pull-based architecture, tensor values are computed only when their chunks are requested from a compute graph that is constructed from input tensors and operators.
This contrasts with a push-based architecture (e.g., implemented by Voreen~\cite{drees2022voreen}), where the compute graph is evaluated in a push-based manner starting from the input nodes, computing full result tensors at each step until a terminal node (e.g., a canvas or a save-to-disk node) is reached.

\begin{figure}[t]
  \begin{python}
import palace as pc
import numpy as np

# Input
time_series_4d = pc.open("path.h5")
raw_vol = time_series_4d[27,:,:,:]

# Tensor processing
kernel = np.array([1.,2.,1.], dtype=np.float32)*0.25
raw_vol = raw_vol.cast(pc.ScalarType.I16)
smooth_vol = raw_vol.separable_conv([kernel]*3)
vol = (smooth_vol - raw_vol).abs()

fov = 30.0
frame_size = [1920, 1200]
tile_size = [512]*2
config = pc.RaycasterConfig()
tf = pc.grey_ramp_tf(min=0.0, max=1.0)
camera_state = pc.CameraState.for_volume(
  vol.metadata, vol.embedding_data, fov)
frame_md = pc.TensorMetaData(frame_size, tile_size)

# Rendering pipeline
proj = camera_state.projection_mat(frame_md.size)
eep = pc.entry_exit_points(vol.metadata,
    vol.embedding_data, frame_md, proj)
lod = vol.single_level_lod()
frame = pc.raycast(lod, eep, config, tf)

# Frame processing
frame = frame.cast(pc.ScalarType.F32.vec(4))
smooth_frame = frame.separable_conv([kernel]*2)
frame = (smooth_frame - frame).abs()
final_frame = frame.cast(pc.ScalarType.U8.vec(4))

# Create runtime
rt = pc.RunTime(ram_storage_size=10<<30,
                vram_storage_size=10<<30)
# Query top left rendered tile
# Only here actual computation happens
top_left_tile = rt.resolve(final_frame, [[0]*3])

  \end{python}
    \caption{
      A Palace processing and rendering example:
      A 4D tensor (volume time series) is loaded and sliced to obtain a 3D tensor (single volume).
      The volume is then processed using a separable convolution and following point-wise operations.
      It is then used as the input for a volume rendering pipeline the result of which can further be processed.
      Using a Palace runtime the created computation graph can then be evaluated to (for example) query and actually compute part of the defined 2D rendering.
    }\label{fig:code}
\end{figure}

\autoref{fig:code} shows a small (but functional) example of a rendering pipeline implemented using the Palace python bindings.
It involves reading a 4D tensor from an HDF5 file, slicing in the time dimension, tensor processing, 3D raycasting and pixel processing of the resulting frame.
Most of the code (up until creating \texttt{final\_frame}) only constructs the computation graph (with \textit{operators} as nodes and input/output relationships as edges) and uses very little computational resources.
Source operators (i.e., operators that do not require other operators as inputs) read tensors from disk, generate tensors procedurally, or provide in-memory NumPy~\cite{harris2020numpy} arrays.
Composition operators include pointwise (inter- and intra-tensor) operations, convolution, slicing, resampling, rendering (volume slice views, volume raycasting, image viewing), and more sophisticated algorithms such as the random walker algorithm for image segmentation~\cite{grady2006random}.
Any node in the graph can be used to query specific chunks from the resulting tensor, save the tensor to disk, or -- in the case of 2D tensors -- render it to a window when building an interactive application.
Only the latter cases involve traversal and computation of the previously virtual results of the computation graph.
An example is the last line of~\autoref{fig:code}, where the top-left tile of the output image is queried and returned as a NumPy array for further in-memory processing or inspection.

\subsection{Architecture Overview}
\label{sec:overview}

The user-defined compute graph (the operator network), which is static for each frame or resolve call, forms the basis of the internal dynamic \textit{task graph}.
Chunk requests to an operator are handled by spawning \textit{tasks}.
Tasks execute the operator's code for a batch of requests, fulfilling them by providing the requested data in GPU memory, main memory (RAM), or on disk.
To fulfill requests, tasks can create requests to global resources or to other operators (specifically, the input operators of the originating operator).
Available request types include
\begin{itemize}
  \item chunk requests to other operators (in particular, to those operators that are the input of the operator that they originate from), which in turn may spawn other tasks, transitively growing the task graph,
  \item spawning and waiting for tasks on the cpu compute thread pool,
  \item GPU related such as command buffer submission, command buffer completion, pipeline barrier submission
  \item allocation and waiting for memory reclamation (see \autoref{sec:memory}),
  \item and waiting for other internal events.
\end{itemize}
Importantly, requests are not necessarily scheduled to be fulfilled by a task immediately.
For example, barrier submissions can (and often are) delayed so multiple tasks can jointly wait on the same compatible barrier, avoiding superfluous synchronization and improving GPU resource utilization
This is enabled by tracking priority per task: allocation and data-transfer operations are assigned higher priority than chunk requests, while barrier handling and memory reclamation receive lower priority.
Besides task class, priority is influenced by the task's progress (assuming tasks that have made more progress are closer to reaching their goal and will free resources sooner) and by the task's depth in the compute graph.

Additionally, the scheduling is influenced by the maximum number of requests per task and the maximum number of active tasks per operator.
Restricting task concurrency is important because deep, wide compute graphs (e.g., on-the-fly LOD construction with many convolution and downsampling stages) can otherwise lead to memory deadlock, where tasks have allocated memory but cannot proceed without additional allocations.
Other strategies to avoid deadlock exist~\cite{weikum2002transactional}, but we do not implement them here for two reasons: (1) computing the required memory to fulfill a task is infeasible because it may include memory allocated by downstream tasks, and (2) tasks cannot currently be aborted.
However, in practice, controlling concurrency via these parameters already avoids deadlocks while still enabling broad and efficient compute resource utilization.

In the implementation, tasks are represented by Rust~\cite{williams2024rust} futures that are executed concurrently, but not in parallel.
The runtime iteratively selects the task with the highest priority from the set of runnable tasks in the task graph (i.e., tasks not currently waiting on other requests).
When a task finishes or reaches an await point, fulfilled and newly created requests are used to update the task graph, and the next task is selected for execution.
This model allows sharing of resources between tasks and modification of management data (e.g., the task graph) without thread synchronization, while still permitting concurrent resource utilization.
Notably, tasks delegate compute-heavy work and blocking I/O operations to the thread pool, preventing the managing thread from becoming a bottleneck.

\begin{figure*}[t]
    \includegraphics[width=\textwidth]{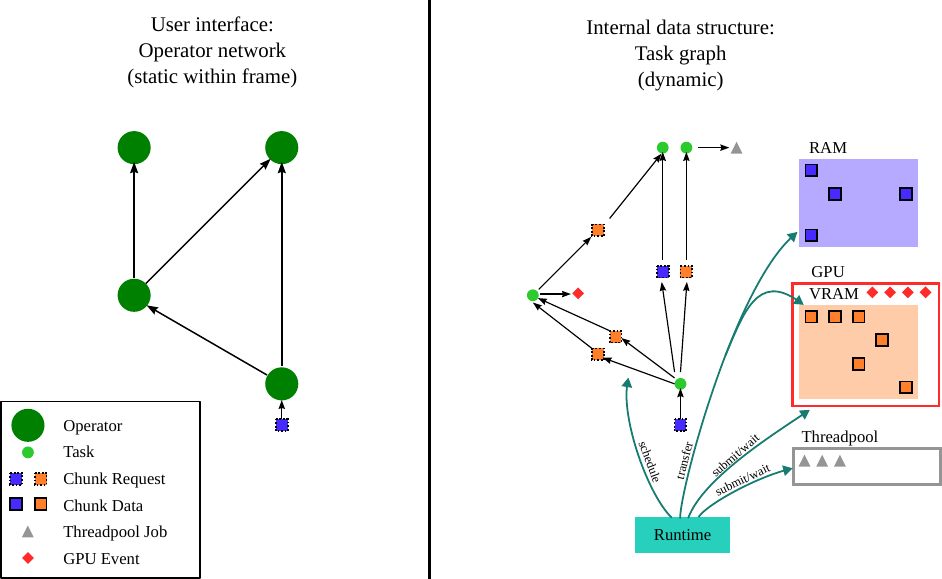}
    \caption{
      Overview of operator network and resulting dynamic request and task network.
      While the users' view is an compute graph (the operator network), static for each resolve-call to the library, during fulfillment of the request, tasks and transitive requests are tracked using the task graph structure.
    }\label{fig:overview}
\end{figure*}

\autoref{fig:overview} shows an illustrative example of the user's view of an operator network and the resulting internal task graph and resource management.

\subsection{Memory}
\label{sec:memory}
All chunks are identified by a corresponding 128-bit ID, which is automatically derived from the operator's ID (recursively determined by its parameters) and the chunk position.
Each store (RAM, VRAM, and disk) tracks the size and memory location of all currently managed chunks within bounded memory regions that are allocated when initializing the runtime.
This ensures Palace does not accidentally exhaust host resources and can manage eviction of unused chunks itself.
For a given chunk ID, the system can determine whether the chunk is already present in the requested location (e.g., GPU memory).
A request completes immediately if the requested chunk is already available in the requested location.
If it is available in another location, a transfer task is automatically dispatched to copy the data into the requested location.
Only if the data is not available in any location, a task for the corresponding operator is run to compute the chunk data.

The concept of locations also enables use of multiple GPUs.
Tasks are executed with a location hint, which is used to select a GPU for subsequent operations.
A node in the operator graph can be annotated to override location hints, enabling distribution of work across multiple GPUs.

For chunks in GPU memory, additionally the visibility of the data is tracked, i.e. which command buffer epoch can read the current version of the data.
If a requested chunk is present in GPU memory but not visible to the destination (per the destination properties), a memory-barrier request is submitted automatically.
The requesting task continues only when the data is ready to be read (e.g., by the compute kernel).

Managing fixed but non-uniformly sized memory regions for chunk storage is more challenging than in typical rendering systems (e.g., Sarton et al.~\cite{sarton2020gpuooc}) for two reasons:
1) Allocations of non-uniformly sized regions rule out simple 1-to-1 memory frame replacement (as in~\cite{sarton2020gpuooc}).
2) There is no explicit synchronization between tasks, so repurposing memory could lead to memory-safety issues.
We address this by using a general-purpose memory allocator for both CPU and GPU memory and invoking a garbage-collection run when allocation for new chunk data fails.
Data eligible for deletion is managed in an LRU queue.
Each garbage-collection run pops items from the queue and frees the corresponding memory until a collection target is reached (typically 10\% of total capacity).
For GPU memory, the process also stops if it encounters an item whose command-buffer epoch has not yet completed (tracked by an epoch counter).

Workloads that frequently allocate and free auxiliary buffers may cause memory fragmentation and incur significant allocator overhead.
To mitigate this, we add a size-bucket cache that collects freed allocations and returns them for reuse instead of creating new allocations through the allocator.
The per-size buckets grow and shrink automatically; if garbage collection cannot free sufficient memory, the cache is flushed.
Allocation sizes are quantized (rounded up) to the nearest $\frac{1}{256}$th increment of the requested size to group similar allocations into the same bucket while avoiding large size overheads.


\subsection{LOD Pyramids}
An LOD (level-of-detail) representation of a high-resolution tensor is especially useful for rendering and for processing operations~\cite{kaufmann1983scalespace,gauch1999watershed,drees2022octree}.
In Palace, an LOD pyramid is an array of tensors~\cite{hadwiger2012pth} together with embedding information subject to a few constraints.
Embedding information includes element spacing (used during rendering).
The spacing should be chosen so that the physical size (spacing multiplied by tensor size) is uniform across all pyramid levels.
The full-resolution tensor (level 0) is the first entry; tensor dimensions decrease and spacing increases for successive levels.

Typically, coarser levels are created by smoothing and downsampling higher-resolution levels.
Palace includes a tool for accelerated offline LOD creation, but pyramids can also be constructed at runtime.
In that case, each level can be described by smoothing and downsampling operations applied to the previous level.
For small volumes, or if a sufficiently large disk cache is configured (at the cost of higher latency), the LOD pyramid can be created on the fly.

\subsection{Page Table Hierarchy}
\label{sec:pth}
For some operators (such as pointwise operations or convolutions), it is straightforward to determine and fetch required chunks in a task before dispatching a GPU compute kernel that directly computes the result.
In other cases, the relationship between parameters and required input chunks can be as hard to determine as the compute step itself.
A typical example is a volume raycaster, which needs to fetch bricks intersecting rays cast through the volume.
The same issue arises for other operators such as a slice or image viewer, or for resampling after a geometric coordinate transformation.
Querying the memory location of every encountered chunk from the CPU would incur a noticeable performance overhead.
Instead, we implement a page-table hierarchy~\cite{hadwiger2012pth} (also used by Sarton et al.~\cite{sarton2020gpuooc}) so that an entire tensor -- or the sparse set of loaded chunks -- can be traversed directly from GPU code.
However, there are key differences from the architecture and implementation of Sarton et al.~\cite{sarton2020gpuooc}:
1) There is one page directory per tensor instead of a multi-resolution page directory.
For operators that require multiple-resolution versions of the same tensor, an array of page directories is passed to the compute kernel and indexed by level.
2) Insertion and removal of pages from the hierarchy are executed in parallel on the GPU, using atomic operations.
3) There is a separate CPU-managed LRU queue for chunk indices and page-table pages; entries are removed from this queue during garbage collection.
Freed page-table pages are unreferenced in the hierarchy, and their allocation can be reused once the command-buffer epoch has expired.
Chunks freed from the page table have their active reference counts decremented, allowing them to return to the LRU queue described in \autoref{sec:memory}.
4) Compute kernels note the use of chunks and page table pages via a hash table, similar to the query buffer by Fogal et al.~\cite{fogal2013hashquery}.
The table is downloaded and read to update the CPU-side LRU table after each kernel invocation.
The hash table does not guarantee reporting of uses (because it is fixed-size and uses bounded linear probing), so the LRU strategy is approximate.
In practice, comparatively small tables (e.g., 2048 entries for a $512\times512$ rendering tile) almost never produce missed reports.
Compared to a one-entry-per-chunk buffer~\cite{sarton2020gpuooc}, this approach has the advantage of a small, constant memory and bandwidth footprint, allowing extremely large volumes to be rendered (see \autoref{sec:zettabyte}).


\subsection{Reducing memory transfers}
For tensor processing on GPUs, the bottleneck often lies in data transfers rather than computation~\cite{eklund2013processinggpu,lutz2020pump}.
This is especially true for larger-than-main-memory tensors that may require writing intermediate results to disk, but also for smaller tensors that fit in RAM/VRAM but not in caches.
Palace's pull-based architecture mitigates this by computing only the intermediate results required for the desired output, thereby reducing overall memory usage.

Additionally, the memory system supports replacing tensor chunks within tasks, enabling operators to operate in-place when a chunk is not concurrently read by another task.
This further alleviates memory pressure.

Finally, pointwise operators (computing a tensor element for the element with the same position in other tensors) can be fused into one, performing all computations in a single compute kernel, avoiding intermediate reads/writes to VRAM.
This happens transparently to the user.
For example, the pixelwise absolute-difference operation and the subsequent type conversion in the frame processing shown in \autoref{fig:code} are performed in a single compute kernel.
Fusing of other operators is not trivial and not implemented so far, but at least point-wise operations with other operators in input/output is possible and will be implemented in the future.
Fusing other kinds of operators is nontrivial and not yet implemented.
However, fusing pointwise operations with adjacent operators in the input/output chain is possible and planned for future work.

\subsection{Operator State}
\label{sec:state}
In general, and in particular from the user's perspective, operators are stateless.
However, internally, most operators keep some state, mostly for performance reasons.
An example is compute kernel programs, which are compiled lazily upon first use and retrieved from a state cache keyed by an ID computed from compilation parameters (e.g., source and defined constants).
Additionally, tasks can opt in to use the state cache to pass and reuse information between invocations with identical parameters, supporting progressive rendering (see~\autoref{sec:rendering}).
Cached state is allocated in the store and is automatically cleaned up after a period of disuse.

\subsection{Rendering}
\label{sec:rendering}
Palace's rendering processors use a hash table~\cite{fogal2013hashquery} for passing brick requests from compute kernels to the CPU-side code.
As with the use tables described in \autoref{sec:pth}, kernels insert the linear IDs of required chunks (and potentially additional information such as the LOD level) into a fixed-size hash table using linear probing
The buffer is read on the CPU side; required chunks are then requested and inserted into the hash table before the kernel is executed again.

Palace also supports \textit{progressive rendering}, supported by bookkeeping in the memory stores.
This is used in rendering processors such as the volume raycaster, slice viewer, and image viewer.
A task can choose to mark generated chunks as \textit{preview} (rather than the \textit{final} result).
Notably, results of downstream processors are also transitively marked as \textit{preview} automatically.
Subsequent queries to operators that generated a preview (even with the same parameters) will not return the previous result; instead they spawn a new task that can, using the state cache (see \autoref{sec:state}), resume computation and produce a progressively more accurate result.
The raycaster first renders, with low latency, a low-resolution version of the full view (using coarser LOD levels), and then iteratively refines it to the full-resolution version.

It should be noted that downloading query and use buffers, submitting barriers, querying new chunks, and inserting addresses into the page-table hierarchy are all asynchronous operations within the progressive rendering loop.
To fully utilize GPU resources, rendering is performed in tasks that compute tiles of a specified size (e.g., $512\times512$ pixels).
Tasks execute concurrently so that, for example, one task can enqueue new graphics commands while another waits for the query-buffer download to finish.
Additionally, a progressive rendering loop begins by reading the cached query buffer filled by a previous preview task, further increasing concurrency and avoiding GPU stalls.
\begin{figure*}[t]
    \includegraphics[width=\textwidth]{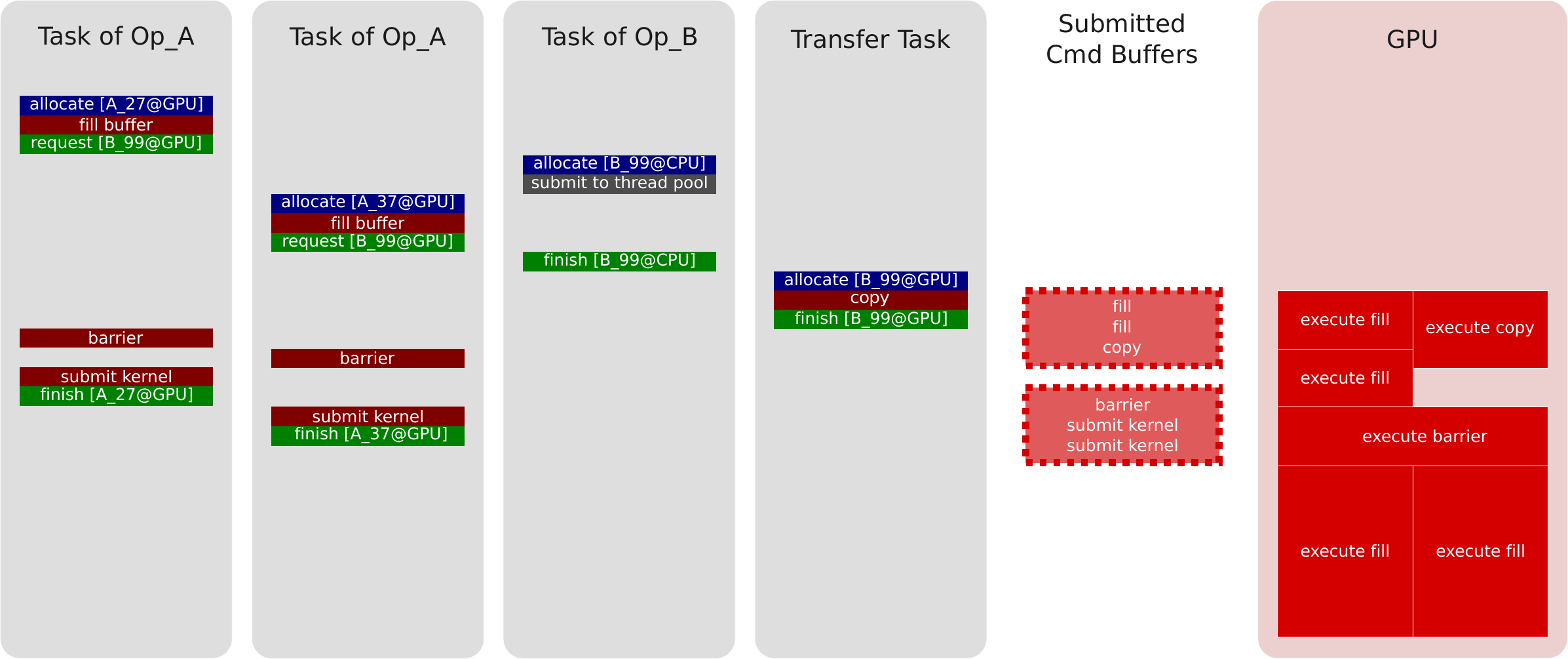}
    \caption{
      Illustration of concurrent execution in Palace.
    }\label{fig:concurrency}
\end{figure*}
The concurrent execution of tasks and GPU kernels is illustrated in \autoref{fig:concurrency}.

Finally, rendering can be optionally accelerated by empty-space skipping~\cite{hadwiger2018skipping}.
In Palace, this is implemented using a \textit{const chunk table}: a tensor whose elements correspond to chunks of the referenced (potentially sparsely populated) tensor.
A sentinel value indicates that the corresponding chunk is not uniform, so the full-resolution tensor must be queried.
Other values encode the uniform value of all elements in the corresponding chunk.
Const chunk tables can themselves be chunked.
They can be computed offline using tools provided by Palace or generated on the fly, similarly to on-the-fly LOD creation.

\section{Quantitative Evaluation}
This section presents two experiments comparing Palace with state-of-the-art systems in volume rendering and processing.

\subsection{Volume raycasting}
\begin{figure*}[t]
  \begin{subfigure}[b]{0.196\textwidth}
    \centering\includegraphics[width=\textwidth]{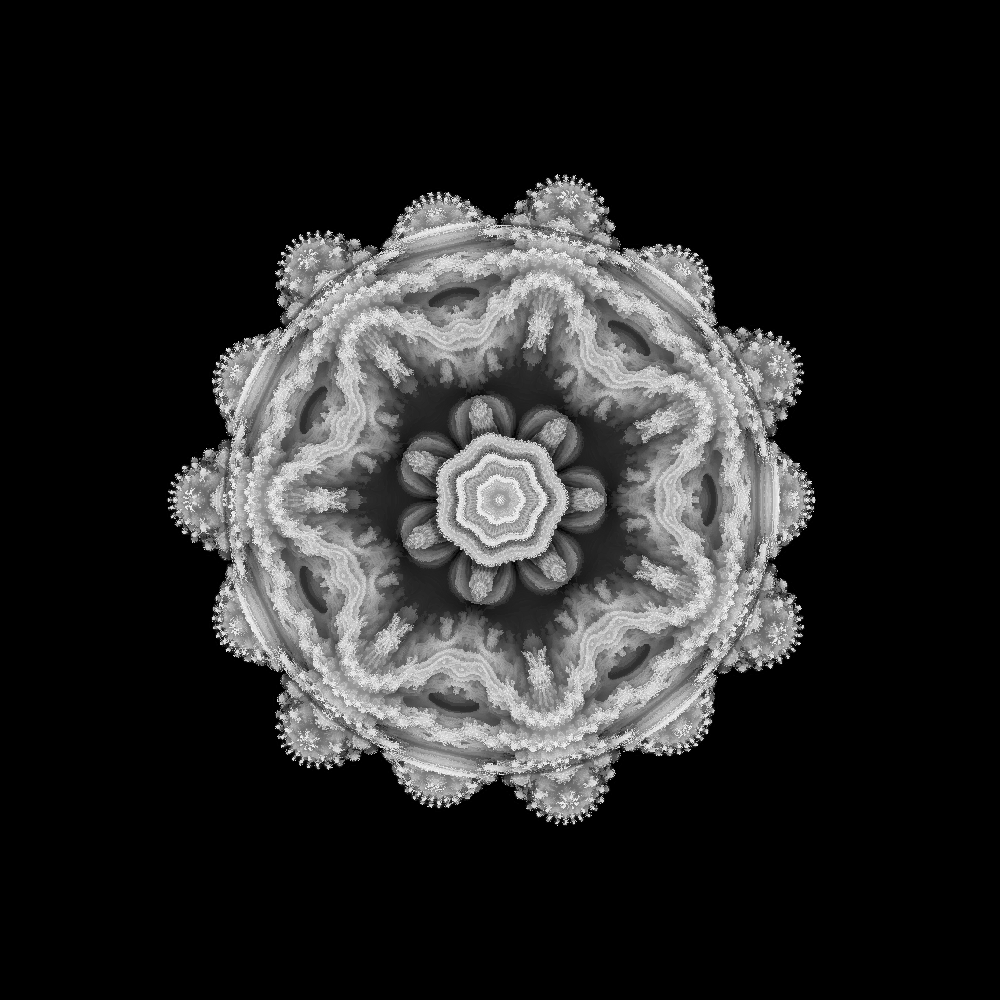}
    \caption{Mandelbulb far (DVR)}
  \end{subfigure}
  \begin{subfigure}[b]{0.196\textwidth}
    \centering\includegraphics[width=\textwidth]{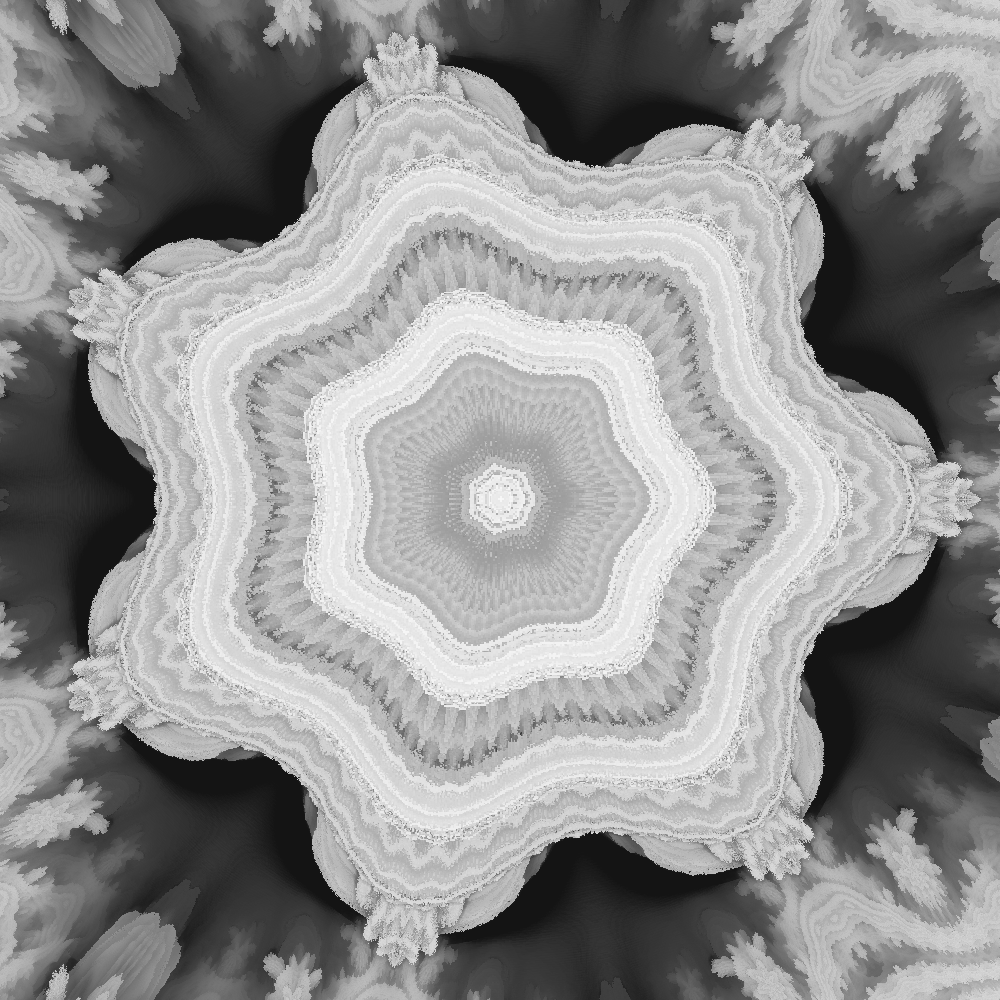}
    \caption{Mandelbulb near (DVR)}
  \end{subfigure}
  \begin{subfigure}[b]{0.196\textwidth}
    \centering\includegraphics[width=\textwidth]{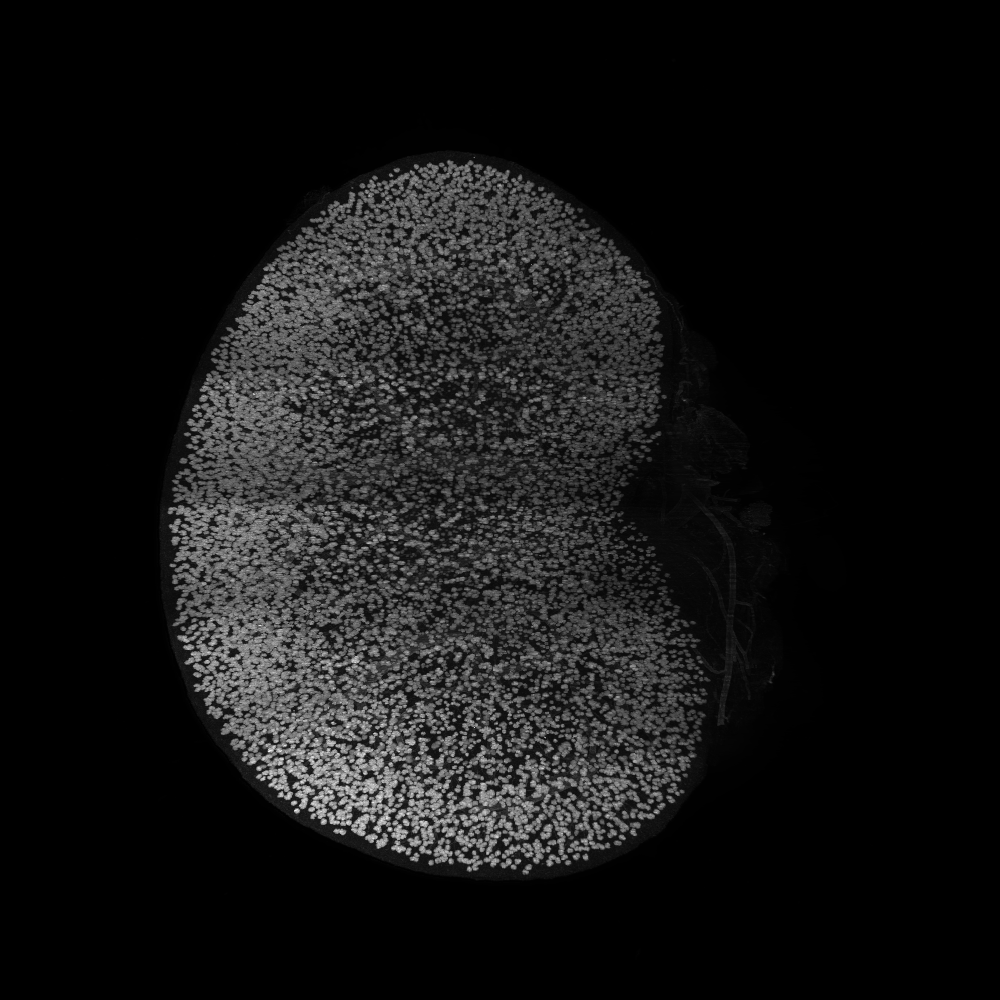}
    \caption{Kidney far (MOP)}
  \end{subfigure}
  \begin{subfigure}[b]{0.196\textwidth}
    \centering\includegraphics[width=\textwidth]{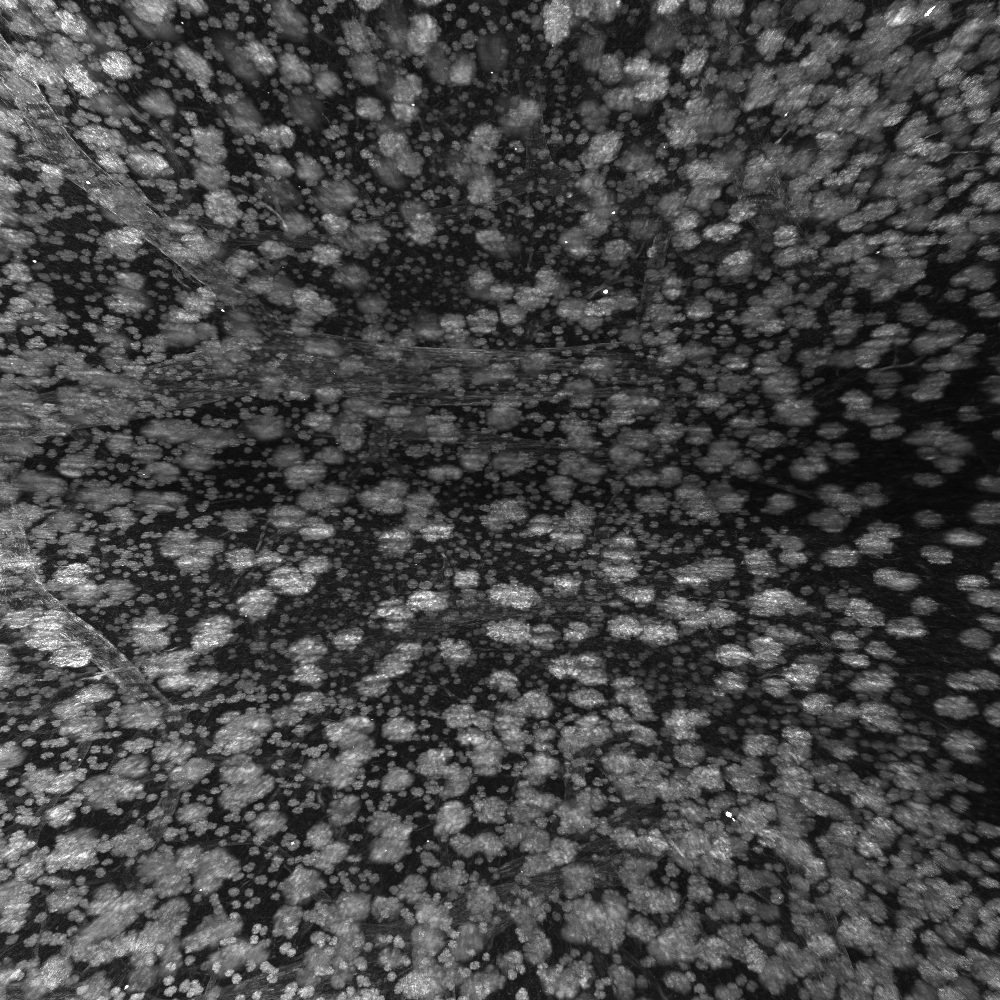}
    \caption{Kidney near (MOP)}
  \end{subfigure}
  \begin{subfigure}[b]{0.196\textwidth}
    \centering\includegraphics[width=\textwidth]{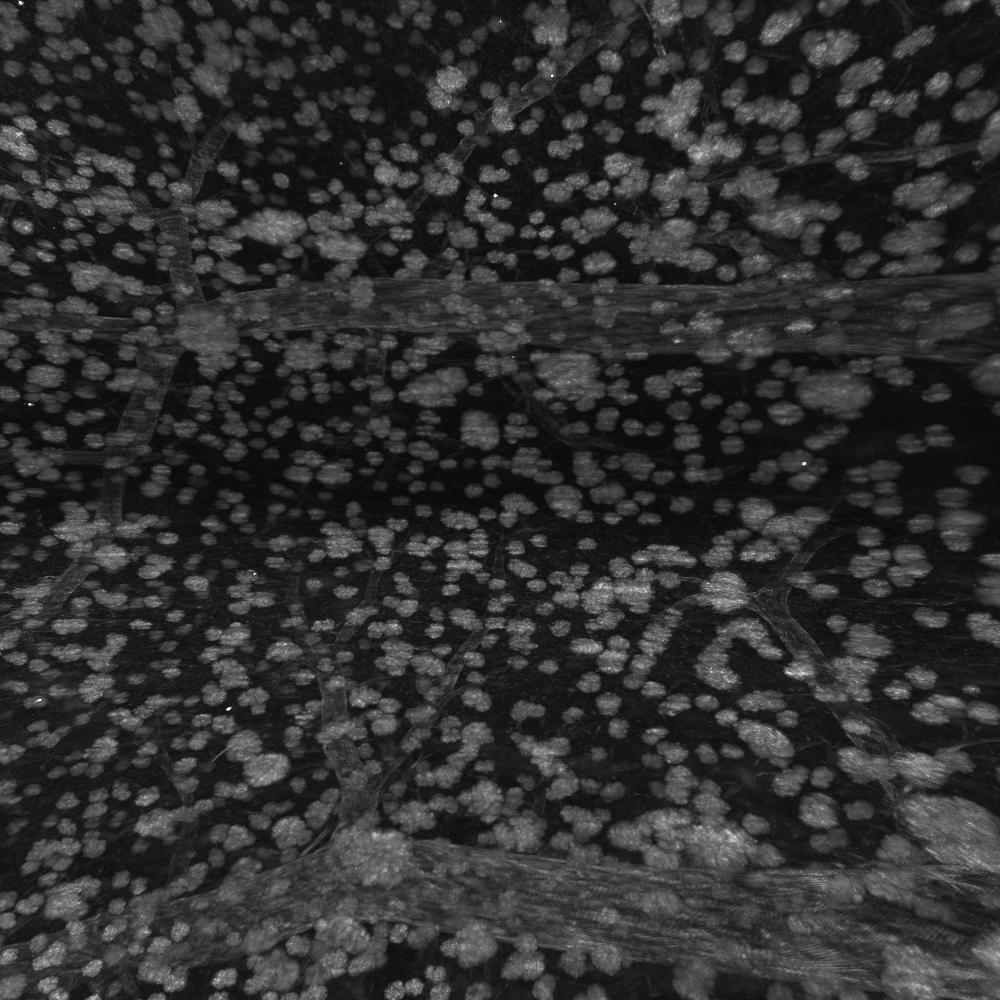}
    \caption{Kidney inside (MOP)}
  \end{subfigure}
  \caption{
    Benchmark views rendered as part of the evaluation as reported in \autoref{tab:bench_raycast}.
    Notably, the kidney appears squashed due to~\cite{sarton2020gpuooc} not supporting anisotropic voxel spacing which has been mimicked in Palace in this experiment for a fair comparison.
  }\label{fig:bench_raycast_views}
\end{figure*}

To evaluate out-of-core raycasting performance, we measured the time to render a complete $1000\times1000$-pixel frame with a cold disk cache.
The experiments were conducted on an AMD Ryzen Threadripper 1920X machine with 64 GB RAM, a GeForce RTX 2080, and a Samsung 990 Pro solid-state drive.
Palace's performance is compared with the state-of-the-art rendering system of Sarton et al.~\cite{sarton2020gpuooc}.
The test setup uses two datasets in different viewing configurations.
The \textit{Mandelbulb} volume was also used in experiments by Sarton et al.~\cite{sarton2020gpuooc} and is rendered using direct volume rendering (DVR).
The \textit{Kidney} dataset is a light-sheet microscopy scan of a murine liver with capillary-level resolution~\cite{kirschnick2021rapid}, rendered using maximum-opacity projection (MOP).
Both datasets are chunked into $64^3$-sized bricks and LOD-pyramid data structures are created offline.
More details are listed in \autoref{tab:bench_datasets}.
\begin{table}
  \begin{center}
    \begin{tabular}{l l l l l}
      Name & Dimensions & Type & Size & Comp.\\\hline
      Mandelbulb & $4352\times4352\times4352$ & float32 & 330GB & DVR\\
      Kidney & $1634\times12723\times9070$ & uint16   & 377GB & MOP
  \end{tabular}
  \end{center}
  \caption{
    Datasets used for benchmarking raycasting performance.
  }\label{tab:bench_datasets}
\end{table}
For \textit{Mandelbulb}, two views -- \textit{far} and \textit{near} -- are used
For \textit{Kidney}, three rendering views \textit{far}, \textit{near} and \textit{inside} are compared.
The resulting images are shown in \autoref{fig:bench_raycast_views}.

Measured run times are shown in~\autoref{tab:bench_raycast}.
\begin{table}
  \begin{center}
    \begin{tabular}{l l l l}
      & Sarton~\cite{sarton2020gpuooc} & Palace (no ES) & Palace (ES)\\\hline
      Mandelbulb far  & 1.60s$\pm$0.0023s & 1.38s$\pm$0.0024s    & 0.98s$\pm$0.0063s\\
      Mandelbulb near & 3.47s$\pm$0.0085s & 1.48s$\pm$0.0053s    & 1.29s$\pm$0.019s\\
      Kidney far     & 22.2s$\pm$0.27s   & 1.66s$\pm$0.015s     & 1.36s$\pm$0.0094s\\
      Kidney near    & 200s $\pm$0.96s    & 7.70s$\pm$0.13s      & 6.67s$\pm$0.087s\\
      Kidney inside  & 110s $\pm$0.97s    & 5.17s$\pm$0.045s     & 4.61s$\pm$0.066s\\
  \end{tabular}
  \end{center}
  \caption{
    Raycasting benchmark rendering times required for full view of the datasets and configurations shown in \autoref{fig:bench_raycast_views}.
    For Palace, results with (ES) and without (no ES) empty brick skipping are reported.
  }\label{tab:bench_raycast}
\end{table}
For Palace, times with and without empty brick skipping are reported.
The results reported for Sarton et al.~\cite{sarton2020gpuooc} include empty-brick skipping.
It should be noted that for a fair comparison, the code by Sarton et al.\ was modified in some aspects, all of which \textit{improved} its rendering performance.
See \autoref{sec:sarton} for details.
All experiments were repeated three times; we report the mean and standard deviation.

For direct volume rendering of the Mandelbulb dataset, both systems show times of the same order of magnitude (a few seconds); even without empty-brick skipping, Palace is faster.
For \textit{Mandelbulb near}, Palace is roughly twice as fast.
Enabling empty-brick skipping improves performance further by about 15-40\%.

For the Kidney dataset maximum-opacity projection, Palace requires two to three orders of magnitude less time than Sarton et al.'s system.
In the far view, Sarton et al.'s system benefits from rendering a lower LOD level, but still requires more time than DVR because rays traverse the whole volume.
For close-up views requiring high brick-loading throughput, Sarton et al.'s system can take more than a minute to render a complete frame.
A potential reason for this is the serialized rendering-request loop (not loading bricks while a frame is rendered) and sequential updates to the page table hierarchy.

Palace, by contrast, benefits from interleaved rendering and brick loading across multiple raycaster tasks (one per output tile), which are executed concurrently, together with parallel page-table updates.
As a result, bricks are read at speeds of $\approx$ 2.5 GB/s -- close to the PCIe 3.0 SSD throughput ($\approx$ 3.9 GB/s) -- and rendering times remain on the order of seconds.
This also explains why Palace benefits less from empty-brick skipping: because brick reads are already near the device throughput, skipping empty bricks yields smaller gains.

\subsection{Hierarchical Random Walker Segmentation}

For the evaluation of an out-of-core processing pipeline, we chose the hierarchical random walker segmentation algorithm~\cite{drees2022octree} of which the original implementation in Voreen~\cite{drees2022voreen} has been ported to Palace.
Based on the original paper~\cite{drees2022octree}, we evaluate on a synthetic cell dataset with $2000^3$ voxels and a total size of 16 GB.
We applied the mean-filter-based weight function of Bian et al.~\cite{ang2016statistical}.
Experiments were conducted on an AMD Ryzen 9 5900X machine with 32 GB RAM, an AMD RX 6700XT, and a Samsung 980 Pro solid-state drive.

A notable difference in the implementation is that due to its push-based architecture, Voreen always computes the whole volume while Palace's pull-based architecture allows lazy computation of only those bricks that are visible to the user.

The results are shown in \autoref{tab:bench_rw}; the table reports the full-volume computation time (for example, to store the resulting volume to disk) for both Voreen and Palace.
Additionally, for Palace we report the compute time for a quad-view rendering of an $800\times600$ window (similar to \autoref{fig:4drw}).
\begin{table}
  \begin{center}
    \begin{tabular}{l l l}
      Voreen\cite{drees2022voreen} & Palace (full volume) & Palace (rendered) \\\hline
      98.47s $\pm$0.78s            & 41.47s $\pm$0.69s & 11.71s $\pm$0.37s\\
  \end{tabular}
  \end{center}
  \caption{
    Comparison of hierarchical random walker compute times on synthetic cell dataset~\cite{drees2022octree}.
  }\label{tab:bench_rw}


\end{table}
For the full volume, Palace requires less than half the computation time of Voreen (41.47 s vs. 98.47 s)."
The rendered view (including rendering time but omitting computation of bricks not shown, e.g., finer LOD levels) completes in 11.71 s."
It should be noted, however, that Voreen's implementation includes an optimization for interactive updates which has not yet been implemented in Palace and is not evaluated here.

\section{Use Case Demonstrations}

In addition to the quantitative evaluation, this section presents use cases for Palace: large 2D image viewers, higher-dimensional image processing, and extremely large-volume visualization.

\subsection{Slide Image Rendering}

\begin{figure}[t]
    \includegraphics[width=\columnwidth]{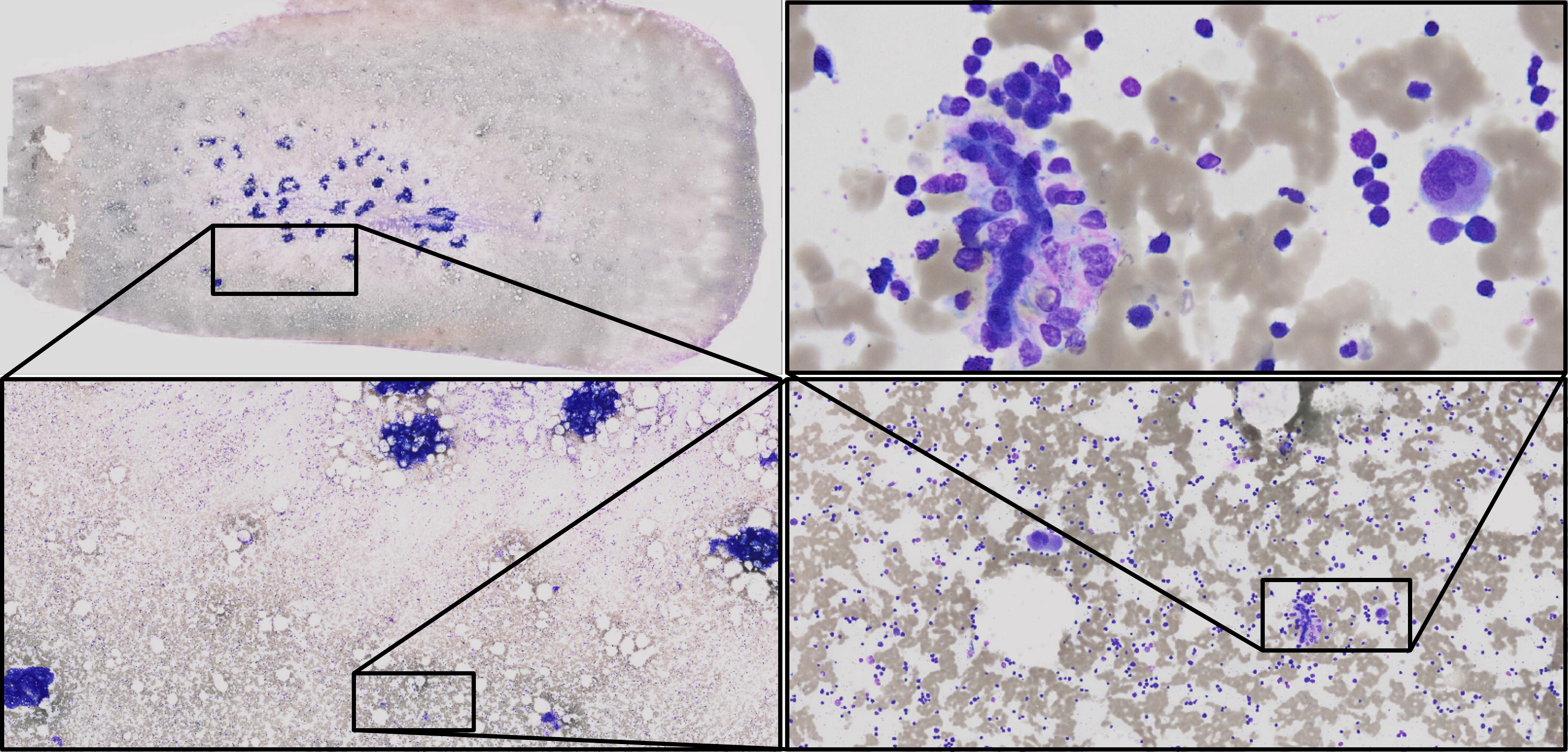}
    \caption{
      Four zoom levels rendered with Palace's image viewer in a whole slide image from a bone marrow smear dataset~\cite{kockwelp2022cvpr,kockwelp2024deep}.
    }\label{fig:slide}
\end{figure}

Palace can also be used to process and visualize 2D datasets.
To demonstrate this, we visualized a whole-slide image from a bone-marrow-smear dataset~\cite{kockwelp2022cvpr,kockwelp2024deep} consisting of $134772\times281817$ RGB pixels and a total size of 114 GB.
Palace's image viewer uses the LOD pyramid to seamlessly switch between resolution levels during zoom interactions.
A composite image with four zoom levels is shown in~\autoref{fig:slide}.

\subsection{4D Random Walker Segmentation}

\begin{figure}[t]
    \includegraphics[width=\columnwidth]{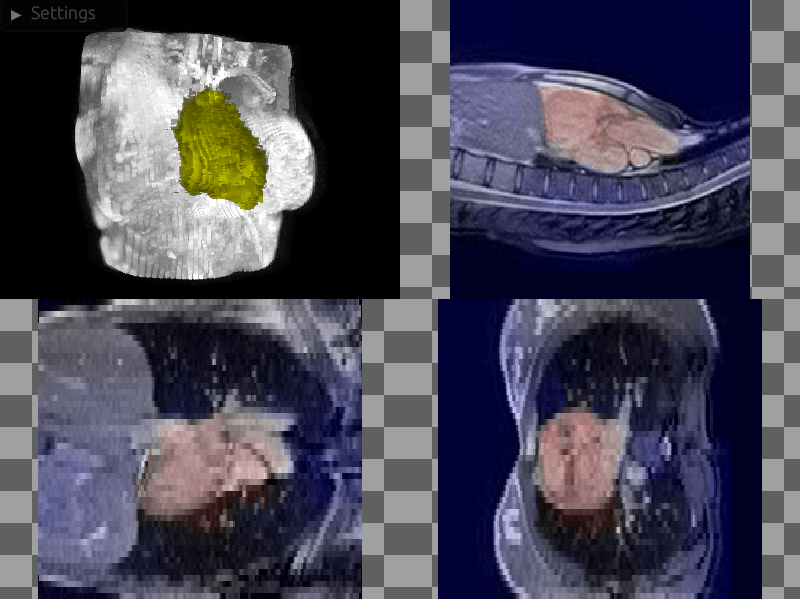}
    \caption{
      Screenshot of example application segmenting the heart in a volumetric time series dataset from seeds in time points 0 and 199, showing segmentation in time step 100.
    }\label{fig:4drw}
\end{figure}

To show flexibility in dataset dimensions, we applied the hierarchical random walker to a 4D MRI dataset~\cite{boye2013population}, with $50\times224\times224$ voxels per time step and 200 time steps (interpreted as a 4D tensor).
The dataset was opened in a quad-view example application showing three axis-aligned slices and a raycasting rendering of the selected time point, which can be changed via the on-screen GUI.
We interactively added 10 foreground seeds and 10 background seeds at time point 0, and 4 foreground seeds at time point 199 via the slice views.
The screenshot in \autoref{fig:4drw} shows time point 100, demonstrating that segmentation for intermediate time points can be obtained with the 4D random walker implementation.

\subsection{Raycasting a virtual Zettabyte-scale Volume}
\label{sec:zettabyte}

One benefit of Palace's architecture is that tensors need not be persistently stored on disk to be loaded.
Procedural on-the-fly generation of chunks is also possible.
In combination with dynamic LOD selection, this enables handling of virtually arbitrarily large volumes.
\begin{figure*}[t]
    \includegraphics[width=\textwidth]{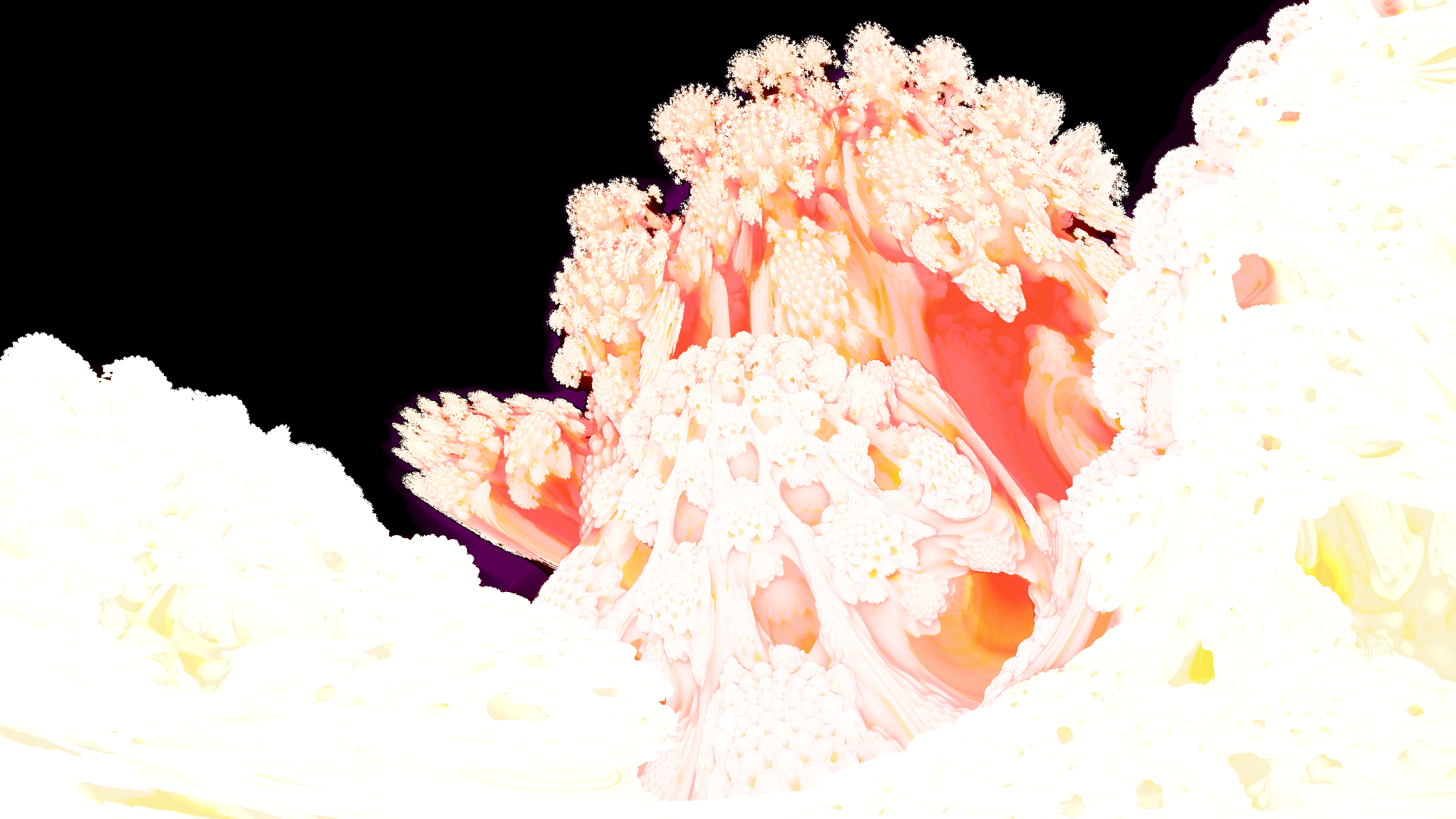}
    \caption{
      Surface view from a procedurally generated, virtual 2ZB mandelbulb volume.
    }\label{fig:zettabyte}
\end{figure*}
\autoref{fig:zettabyte} shows a rendered surface view of a mandelbulb volume of $8000000^3$ voxels rendered with bricks of size $128^3$.
At 4 bytes per voxel this corresponds to a 2 ZB (zettabyte) dataset.
Due to the fixed three-level page table with $2^{16}$ entries per page, this is the upper limit of the current implementation.
Nevertheless, this limit should suffice for real world use cases of non-procedural volumes in the near future.

\section{Limitations}
Most operators are implemented in a dimension-agnostic fashion.
For this reason, the implementation intentionally does not use the sparse-image features provided by Vulkan~\cite{khronos2025vulkan}; instead, each chunk is stored in a separate buffer.
As a result, specialized sampling hardware on the GPU may not be fully utilized.
This may reduce rendering performance -- a trade-off made for greater flexibility.

As noted in \autoref{sec:overview}, deep compute graphs can lead to deadlock.
This can usually be avoided by reducing concurrency via configuration parameters or, preferably, by increasing the memory resources available to Palace.

\section{Conclusion}

We presented Palace, a cross-platform library for interactive, GPU-accelerated, out-of-core tensor processing and visualization.
Through efficient implementations of established methods (such as GPU-side page-table directories) and a novel asynchronous concurrent architecture, Palace surpasses state-of-the-art approaches in the two tested scenarios: volume rendering and hierarchical random-walker segmentation.
We demonstrated applicability across multiple domains and dimensions using case studies.
This is enabled by a simple, functional compute-graph interface.

In future work, we will continue to develop Palace as an open-source project, porting and developing more out-of-core algorithms and improving performance -- for example by extending operator fusion.

\section*{Acknowledgments}
We thank Hennes Rave for helpful discussions during the initial conception of Palace's architecture.

\appendix[Summary of changes to Sarton's code]
\label{sec:sarton}
The code of Sarton et al.~\cite{sarton2020gpuooc}, available online\footnote{\url{https://github.com/joSarton/GPU-Out-of-Core-Volume-Data}} was modified to enable comparison with Palace.
The changes include
\begin{itemize}
  \item re-enabling empty-brick optimization,
  \item dynamic LOD selection during rendering, a port of the strategy of Palace, improving performance for volumes larger than a few gigabytes,
  \item saving render state progress between render calls (for rendering views where the working set is larger than VRAM and resulting in a $\approx$ 10x throughput increase due to less time spent raycasting),
  \item terminating the render loop on unmapped bricks, roughly doubling throughput,
  \item adding MOP compositing,
  \item various smaller changes to ease benchmarking as well as technical fixes.
\end{itemize}
It should be stressed that none of the changes decreased rendering performance, but instead caused the required rendering time in the benchmark to either remain constant or decrease.
The full changed source code is also available online\footnote{\url{https://github.com/ftilde/GPU-Out-of-Core-Volume-Data}}.

\bibliography{src}
\bibliographystyle{IEEEtran}

\begin{IEEEbiography}[{\includegraphics[width=1in,height=1.25in,clip,keepaspectratio]{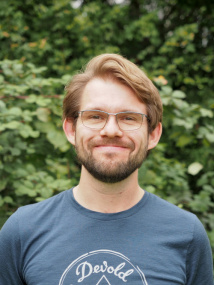}}]{Dominik Drees}
  received a PhD in computer science from the University of M\"{u}nster, Germany, in 2022 and is currently working as a postdoctoral researcher. His research interests are image and volume processing and analysis, resource constrained (esp. out-of-core) computing, as well as nanophotonic neural networks.
\end{IEEEbiography}

\begin{IEEEbiography}[{\includegraphics[width=1in,height=1.25in,clip,keepaspectratio]{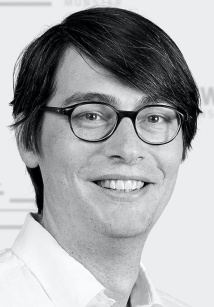}}]{Benjamin Risse}
  has been a professor at the University of M\"{u}nster, heading the Computer Vision \& Machine Learning Systems group since 2018. He studied computer science and received a PhD at the intersection of computer vision, machine learning, and neuroscience. From 2015 to 2017, he was a postdoctoral researcher at the University of Edinburgh. Today, he is the managing director of the Institute for Applied Computer Science, the deputy director of the Institute of Geoinformatics, and an affiliated professor at the Faculty of Mathematics and Computer Science at the University of M\"{u}nster, with a focus on artificial intelligence.
\end{IEEEbiography}

\end{document}